\documentclass[aps,prl,twocolumn,groupedaddress,showkeys,preprintnumbers,nofootinbib]{revtex4-2}

\usepackage[T1]{fontenc}
\usepackage[utf8]{inputenc}
\usepackage{xcolor} 
\colorlet{lightgray}{black!10} 
\colorlet{pencil}{black!25} 
\usepackage{cancel}
\usepackage{dsfont}
\usepackage{tikz}
\usepackage{appendix}
\usetikzlibrary{decorations.pathmorphing,calc}
\usepackage{tikz-feynman}
\tikzfeynmanset{compat=1.1.0}

\usepackage{amsmath}
\usepackage{mathtools}
\usepackage{amssymb}
\usepackage{amstext}
\usepackage{amsfonts}
\usepackage{physics}
\usepackage{slashed}
\DeclareSymbolFontAlphabet{\amsmathbb}{AMSb}
\usepackage{bbold}
\usepackage{leftindex}
\usepackage{bbm}\usepackage{bm}
\usepackage{comment}
\usepackage{framed}

\usepackage[
  colorlinks=true,
  linkcolor=blue,
  citecolor=blue,
  urlcolor=blue
]{hyperref}

\usepackage{comment}

\newcommand{\SU}{\mathrm{SU}}
\newcommand{\U}{\mathrm{U}}

\renewcommand{\braket}[1]{\left\langle{#1}\right\rangle}

\renewcommand{\tr}{\mathrm{tr}}

\newcommand{\de}{\text{d}}

\newcommand{\ie}{\textit{i.e.}~}
\newcommand{\eg}{\textit{e.g.}~}
\newcommand{\cf}{\textit{cf.}~}

\begin{document}

\preprint{DESY-26-139}

\title{Twisted Correlators from Harmonic Superspace}

\author{Dennis le Plat}
\email[]{dennis.leplat@nbi.ku.dk}
\affiliation{HUN-REN Wigner Research Centre for Physics, Konkoly-Thege Mikl\'os u. 29-33, 1121 Budapest, Hungary}
\affiliation{Niels Bohr International Academy, Niels Bohr Institute, Copenhagen University, Blegdamsvej 17, DK-2100 Copenhagen Ø, Denmark}

\author{Torben Skrzypek}
\email[]{torben.skrzypek@desy.de}
%\homepage[]{Your web page}
%\thanks{}
%\altaffiliation{}
\affiliation{Deutsches Elektronen-Synchrotron DESY, Notkestra{\ss}e 85, 22607 Hamburg, Germany}

\begin{abstract}
We establish a harmonic superspace formulation for $\mathcal N=2$ orbifolds of $\mathcal N=4$ super-Yang-Mills (SYM) theory. We account for the orbifold projection by introducing topological defect lines to the planar supergraphs of $\mathcal N=4$ SYM. These twist lines are sourced by twisted sector operators, transparent to vector multiplets but generate orbifold phases when crossed by bifundamental hypermultiplet fields. This formalism establishes a refinement of the orbifold inheritance principle at supergraph level. As a direct application, we compute leading twist-dependent corrections to BPS two- and three-point functions, finding agreement with localisation results in extremal cases and generating new data for non-extremal correlators. We further map these corrections to finite-size effects in the twisted hexagon formalism. This opens a path towards an all-order integrability prescription of orbifold structure constants.
\end{abstract}

\maketitle

Few interacting quantum field theories are as well understood as $\mathcal{N}=4$ Super-Yang-Mills (SYM) theory, thanks to holography \cite{Maldacena:1997re,Witten:1998qj}, integrability \cite{Minahan:2002ve, Beisert:2003tq}, and supersymmetric localisation \cite{Pestun:2007rz}. This control partially extends to orbifold deformations of $\mathcal{N}=4$ SYM \cite{Kachru:1998ys,Lawrence:1998ja}. The resulting integrable quiver gauge theories contain an untwisted sector, which descends from the ``parent'' $\mathcal{N}=4$ SYM and whose observables match their parent counterparts according to the \emph{inheritance principle} \cite{Bershadsky:1998mb, Bershadsky:1998cb}. Additional twisted sectors, however, display genuinely novel behaviour.

In $\mathcal{N}=2$ orbifold theories we have access to the powerful harmonic superspace formalism \cite{Galperin:2001seg} which significantly simplifies perturbative gauge-theory calculations, as we will outline below. In the large-$N$ limit, the computation of correlators boils down to the evaluation of a relatively small number of planar supergraphs. We will show that the only difference between twisted and untwisted correlation functions is the inclusion of topological twist lines, resulting in a refinement of the inheritance principle. We discuss implications for BPS two- and three-point functions, leaving details of the superspace computations to a companion paper \cite{LongPaper}. 

We can compare these results with localisation and integrability where twist-factors serve as organisation principle, suggesting a direct correspondence with supergraphs. We discuss potential extensions to subleading terms in the localisation result and highlight the capacity of harmonic superspace to generate targeted field-theory data for the development and validation of exact techniques, such as an all-order hexagon formalism for orbifold theories. This would establish the same level of calculational control in $\mathcal{N}=2$ orbifold theories as has been achieved for $\mathcal{N}=4$ SYM. 
%Similar diagrammatic matching may be instrumental to uncovering hidden structure in marginal deformations and $\mathcal{N}=1$ orbifold theories.

\section{\texorpdfstring{$\bm{\mathcal N=4}$}
~~SYM in harmonic superspace}

In order to perform the orbifold projection, we first formulate $\mathcal{N}=4$ SYM theory in $\mathcal{N}=2$ harmonic superspace language \cite{Galperin:1984av,Galperin:1984mln,Galperin:1985bj,Galperin:1985va}, where it comprises a vector multiplet with gauge group $\SU(N)$ coupled to an adjoint hypermultiplet. A thorough review of harmonic superspace can be found in \cite{Galperin:2001seg}. In brief, fields on harmonic superspace are functions of four bosonic space-time coordinates $x$, two sets of Grassmann coordinates $\theta^{\pm}, \bar{\theta}^{\pm}$, which are themselves Weyl spinors, and harmonic coordinates $u^\pm$ in the fundamental representation of $\tfrac{\SU(2)}{\U(1)}$. The expansion in the Grassmann coordinates terminates as usual, while the expansion in harmonic coordinates generates an infinite number of auxiliary fields. In this letter we will mostly suppress the $u^\pm$ dependence, referring the reader to the companion paper \cite{LongPaper} for details. We only clarify that indices $\pm$ denote the quotient-$\U(1)$ weight which has to remain homogeneous in all calculations.

The action of $\mathcal{N}=4$ SYM now consists of two terms \cite{Galperin:1985va,Zupnik:1987vm}
\begin{align}
      S_{\mathcal{N}=4}&=S_{V}+S_{H}\,,\label{eq: ActFull}
      \\
      S_{V}&=\frac{1}{4}\int \dd u\, \dd^4 x \,\dd^2\bar\theta^+\,\dd^2\bar\theta^-\, \Tr(\bar W\bar W)\,,\label{eq: ActSYM}\\
      S_{H}&=\frac{1}{2}\int \dd u\, \dd^4 x \,\dd^2\theta^+\,\dd^2\bar\theta^+\, \Tr(q^+_a \mathcal{D}^{++}q^{+a})\label{eq: ActHyp}\,,
\end{align}
where $\bar{W}$ is the field strength associated to the gauge prepotential $V^{++}$ and the analytic superfield $q^{+}$ captures the hypermultiplet. The homogeneous superspace derivative $D^{++}=u^{+}\frac{\partial}{\partial u^{- }}+\dots$ is promoted to a covariant derivative 
\begin{equation}
    \mathcal{D}^{++}q^{+a}=D^{++}q^{+a}+ig_{\textrm{YM}}[V^{++},q^{+a}]\,,
\end{equation}
mediating a minimal coupling to the gauge field. The index $a$ makes manifest an additional rigid $\SU(2)$ Pauli-G\"ursey (PG) symmetry exchanging conjugate fields $q^+_a=(q^+,-\tilde{q}^+)$, which enhances the R-symmetry $\SU(2)\times \U(1)\to \SU(4)$ as required for ${\cal N} = 4 $ SYM.

In the large-$N$ limit we keep the rescaled coupling $g:=g_{\textrm{YM}}\sqrt{N}/4\pi$ finite. A calculation of correlators boils down to the evaluation of planar supergraphs, which involve propagators for matter $q^+$ and gauge $V^{++}$ superfields, a cubic vertex $q^{+}_1\tilde q^{+}_2V^{++}$ as well as an infinite number of $V^{++}$ self-coupling vertices: 
\begin{align*}
&\begin{array}{c c}
\begin{tikzpicture}[baseline=(b.base)]
\begin{feynman}
\feynmandiagram [horizontal=a to b] {
  a -- [fermion] b,
};
\end{feynman}
\end{tikzpicture}
&\sim\frac{i}{p^2}(D^+_1)^4(D^+_2)^4\delta^8(\theta_1-\theta_2)\,,
\end{array}\\
&\begin{array}{c c}
\begin{tikzpicture}[baseline=(b.base)]
\begin{feynman}
\feynmandiagram [horizontal=a to b] {
  a -- [photon] b,
};
\end{feynman}
\end{tikzpicture}
&\sim\frac{2i}{k^2}(D^+_1)^4\delta^8(\theta_1-\theta_2)\,,
\end{array}\\
&\begin{array}{c c}
\begin{tikzpicture}[baseline=(v.base)]
\begin{feynman}
% Define vertices manually
\vertex (v) at (0.5,0);
\vertex (l) at (0,0);
\vertex (r) at (1,0);
\vertex (t) at (0.5,0.42);
% Draw edges
\feynmandiagram [horizontal= (l) to (r)]{
  (l) -- [] (v) -- [fermion] (r),
  (t) -- [photon] (v),
};
\end{feynman}
\end{tikzpicture}
&\sim -i g (2\pi)^4\delta(p_1-p_2-k)\int  \dd^4\theta^+\,,
\end{array}\\
&\begin{array}{c c}
\begin{tikzpicture}[baseline=(v.base)]
\begin{feynman}
\vertex (v) at (0.5,0);
\vertex (b1) at (0,-0.4);
\vertex (b2) at (1,-0.4);
\vertex (t1) at (0,0.4);
\vertex (t2) at (1,0.4);
\feynmandiagram [horizontal= (b1) to (b2)]{
  (b1) -- [photon] (v),
  (t1) -- [photon] (v) -- [photon] (t2),
};
\end{feynman}
% Ellipsis
\fill (v) ++(0.25,-0.10) circle (0.03);
\fill (v) ++(0.20,-0.20) circle (0.03);
\fill (v) ++(0.10,-0.25) circle (0.03);
\end{tikzpicture}
&\sim-\frac{(-ig)^{n-2}}{2n}(2\pi)^4\delta(\sum k_i)\int \dd^8\theta \,,
\end{array}
\end{align*}
where $D^\pm$ denotes super-covariant spinor derivatives and $p,k$ are the momenta of hyper- and vector-multiplet fields, respectively.
In these Feynman rules, we suppressed the gauge structure, which can be introduced using 't Hooft double-line notation \cite{tHooft:1973alw}, as well as ghosts and the harmonic $u^\pm$-space structure, which are discussed in \cite{Galperin:2001seg,LongPaper}. Note furthermore that the hypermultiplet propagators point from $q^{+}$ to their conjugate $\tilde{q}^{+}$ insertions.

\section{\texorpdfstring{$\bm{\mathcal N=2}$}{N=2} orbifolds and twist lines}

For simplicity, we focus on \emph{abelian} $\mathcal{N}=2$ orbifold theories. They are constructed by projecting $\mathcal N=4$ SYM to invariant states under a $\amsmathbb{Z}_M$ automorphism, which acts on the PG symmetry 
\begin{equation}
    \Gamma:~(q^{+},\tilde q^{+})\to (\omega q^{+},\omega^{-1}\tilde q^{+})\,,\qquad \omega=e^{\frac{2\pi i}{M}}\,, 
\end{equation}
and simultaneously by the adjoint action of
\begin{equation}\label{eq: twistop}
    \gamma=\text{diag}(\mathbb{1}_{N},\omega\cdot\,\mathbb{1}_{N},\omega^2\cdot\,\mathbb{1}_{N}\,,\dots\,,\omega^{M-1}\cdot\,\mathbb{1}_{N})
\end{equation}
on $\SU(MN)$ colour space. This breaks the gauge group down to $\SU(N)^M\times \U(1)^{M-1}$. The $\U(1)$-factors break conformality and are removed at the conformal fixed point \cite{Dymarsky:2005nc,
Gadde:2010zi}. The resulting CFT retains $\mathcal{N}=2$ supersymmetry and is thus naturally described by a harmonic superspace action 
\begin{equation}\label{eq: orbact}
    S_{\mathcal{N}=2}=\sum_{I=0}^{M-1} (S_{V,I}+S_{H,I})\,,
\end{equation}
involving $M$ $\SU(N)$ gauge multiplets  $V^{++}_m$ with action \eqref{eq: ActSYM} linked by $M$ hypermultiplets $q_I^+$ with action \eqref{eq: ActHyp}, in which the covariant derivatives now feature an appropriate bifundamental coupling \cite{LongPaper} to $V_I$ and $V_{(I+1)\text{ mod } M}$.

In order to unify the treatment of these $M$ copies, we introduce \emph{quiver traces} which sum over all $\SU(N)^M$ indices and combine the superfields into $M\times M$ matrices \cite{Ideguchi:2004wm}. Vector multiplets are arranged block-diagonally, while hypermultiplets sit on the off-diagonals
\begin{align}
    q^{+}=\begin{pmatrix} 0 &  q^{+}_0& 0& \hspace{-5pt}\dots\\
   0 & 0& q^{+}_1& \\[-5pt]
    \vdots & && \hspace{-5pt}\ddots \end{pmatrix}\,,\,\, \tilde{q}^{+}=\begin{pmatrix} 0 &  0 & \hspace{-5pt}\dots\\
   \tilde{q}^{+}_0 & 0& \\
    0 & \tilde{q}^{+}_1 & \\[-5pt]
   \vdots & &  \hspace{-5pt}\ddots \end{pmatrix}\,.
\end{align}
In this notation the orbifold action \eqref{eq: orbact} reduces to the familiar $\mathcal{N}=4$ SYM action \eqref{eq: ActFull}. Single-trace operators that survive the orbifold projection are simply those on which the quiver trace closes, \ie where the field numbers balance as
\begin{equation}
    \# (q^{+} )-\# (\tilde{q}^{+})\in M\amsmathbb{Z}\,.
\end{equation}

In order to probe the entire Hilbert space of valid single-trace operators, we also need to allow for twisted sectors. These are generated by inserting the twist operator $\gamma^\alpha$ \eqref{eq: twistop} into the trace, where $\alpha\in \amsmathbb{Z}_M$ determines the twist sector. Since $\gamma$ commutes with $V^{++}$ and satisfies
\begin{equation}
    \gamma^\dagger q^+\gamma =\omega q^+\,,\qquad  \gamma^\dagger \tilde q^+\gamma =\omega^{-1} \tilde q^+  \,,
\end{equation}
we can move the twist operator through the single-trace operator in question and furthermore split and merge twist operators as long as we keep track of the resulting phases. This makes the twist operators topological defects on the chain of operators, in keeping with their holographic origin as boundary condition of twisted strings. 

Introducing these additional twist operators in the planar supergraphs at large-$N$ sources twist lines that can be moved freely through gauge propagators but cause a phase shift for intersecting hypermultiplet propagators: 
\begin{align*}
& \begin{array}{c c}
\begin{tikzpicture}[baseline=(c.base)]
\begin{feynman}
  % Define vertices
  \vertex [label=left:$\alpha_1$](c) at (0,0);
  \vertex [label=right: $\alpha_2$](d) at (1,0);
  % Draw diagram
  \diagram* {
    (c) -- [scalar, red]  (d),
  };
\end{feynman}
\end{tikzpicture}
&\sim \delta(\alpha_1+\alpha_2\text{ mod }M)\,,
\end{array}\\
&\begin{array}{c c}
\begin{tikzpicture}[baseline=(v.base)]
\begin{feynman}
\vertex (v) at (0.5,0);
\vertex [label=left:$\alpha_1$] (b1) at (0,-0.4);
\vertex  (b2) at (1,-0.4);
\vertex [label=left:$\alpha_2$](t1) at (0,0.4);
\vertex [label=right:$\alpha_3$](t2) at (1,0.4);
\diagram* {
  (b1) -- [scalar, red] (v),
  (t1) -- [scalar, red] (v) -- [scalar, red] (t2),
};
\fill[red] (v) circle (0.03);
\end{feynman}
% Ellipsis
\fill (v) ++(0.25,-0.10) circle (0.03);
\fill (v) ++(0.20,-0.20) circle (0.03);
\fill (v) ++(0.10,-0.25) circle (0.03);
\end{tikzpicture}
&\sim \delta(\sum \alpha_i \text{ mod }M)\,,
\end{array}\\[-12pt]
&\begin{array}{c c}
\begin{tikzpicture}[baseline=(v.base)]
\begin{feynman}
  % Define vertices
  \vertex (a) at (0,0.4);
  \vertex (b) at (1,-0.4);
  \vertex [label=left:$\alpha_1$](c) at (0,-0.4);
  \vertex [label=right:$\alpha_2$](d) at (1,0.4);
  \vertex (v) at (0.5,0);
  % Draw diagram
  \diagram* {
    (a) -- [] (v) -- [fermion] (b),
    (c) -- [scalar, red] (v) -- [scalar, red] (d),
  };
\end{feynman}
\end{tikzpicture}
&\sim\omega^{\alpha_1} \delta(\alpha_1+\alpha_2\text{ mod }M)\,.
\end{array}
\end{align*}
\vspace{-10pt}

Since planar diagrams mimic the worldsheet of the dual string, these twist lines should be thought of as branch cuts in the worldsheet CFT connecting twisted vertex-operator insertions \cite{Hamidi:1986vh,Martinez:2025jjq}. They furthermore enforce the following selection rule: 

\begin{center}
\emph{A physical correlator of twisted operators can only be non-vanishing if the combined twist vanishes (mod $M$).}
\end{center}

We have thus established a framework in which the only difference between supergraph calculations in $\mathcal{N}=2$ orbifold theories and $\mathcal{N}=4$ SYM is the possible occurrence of topological twist lines. In particular, processes featuring only vector multiplets are completely unperturbed. The same conclusion holds for closed loops of hypermultiplets which do not encircle a net source of twist, since any twist lines can either be pulled out of the loop or confined within:
\begin{equation}
\begin{array}{c@{\qquad}c@{\qquad}c}

% ============================================================
% LEFT: red dashed line passes through the circle
% ============================================================
\begin{tikzpicture}[thick,baseline=-3pt]
\usetikzlibrary{arrows.meta}

\def\R{0.42}

% Slightly slanted red dashed line
\draw[dashed,red,line width=0.8pt]
    (-1.05,-0.08) -- (1.05,0.08);

% Decorated circle exactly in the original style
\draw[line width=1pt]
    (0,0) ++(\R cm,0)
    arc[
        start angle=0,
        end angle=360,
        radius=\R cm
    ]
    node[pos=0.125,sloped,allow upside down]
    {
        \tikz{
            \draw[-{Triangle[length=5pt,width=5pt]}]
            (0,0)--(0.001,0);
        }
    };

\end{tikzpicture}

&

\displaystyle \Rightarrow

&

% ============================================================
% RIGHT: red line goes around the circle
% ============================================================
\begin{tikzpicture}[thick,baseline=-3pt]
\usetikzlibrary{arrows.meta}

\def\R{0.42}
\def\RedR{0.62}

% Angle and half-length of the original red line
\pgfmathsetmacro{\redangle}{atan(0.08/1.05)}
\pgfmathsetmacro{\redlength}{sqrt(1.05*1.05 + 0.08*0.08)}

% Where the straight part starts bending
\def\shoulder{0.80}

% ------------------------------------------------------------
% Red dashed line
%
% Everything is drawn in coordinates rotated by exactly the
% same angle as the original straight red line.
% Thus the semicircle lies entirely on one side of that line.
% ------------------------------------------------------------
\begin{scope}[rotate=\redangle]

\draw[dashed,red,line width=0.8pt]

    % exactly straight and aligned with original line
    (-\redlength,0)
    --
    (-\shoulder,0)

    % small rounded shoulder
    .. controls
       (-0.73,0)
       and
       (-0.66,0.02)
    ..
    (-\RedR,0)

    % exact upper semicircle
    arc[
        start angle=180,
        end angle=0,
        radius=\RedR cm
    ]

    % rounded shoulder back onto the original line
    .. controls
       (0.66,0.02)
       and
       (0.73,0)
    ..
    (\shoulder,0)

    % exactly straight and aligned with original line
    --
    (\redlength,0);

\end{scope}

% ------------------------------------------------------------
% Decorated black circle exactly in your original style
% ------------------------------------------------------------
\draw[line width=1pt]
    (0,0) ++(\R cm,0)
    arc[
        start angle=0,
        end angle=360,
        radius=\R cm
    ]
    node[pos=0.125,sloped,allow upside down]
    {
        \tikz{
            \draw[-{Triangle[length=5pt,width=5pt]}]
            (0,0)--(0.001,0);
        }
    };

\end{tikzpicture}

\end{array}
\end{equation}
This observation constitutes a \textbf{refinement of the orbifold inheritance principle}, which states that untwisted quantities in orbifold theories descend from $\mathcal{N}=4$ SYM. We can now claim similar inheritance for individual supergraphs in which the twist lines can be pulled away from all hypermultiplet propagators.

\section{Twisted BPS correlators}

This refined inheritance principle has particularly strong implications for twisted Coulomb-branch operators, suggesting that twist-dependent corrections to their correlators are highly suppressed. 

To demonstrate this, consider the twisted single-trace BPS operators built from the vector-multiplet scalars $\phi_I$: 
\begin{equation}
    T_{L}^{(\alpha)}(x) =\frac{1}{\sqrt M}\sum_{I=0}^{M-1}    \omega^{\alpha I} \, \tr\,\phi_I^L(x)\,, \quad \alpha\in\amsmathbb{Z}_M\,,
    \label{eq:twisted-operator}
\end{equation}
which in our superspace formalism are the lowest components of the twisted operators $\tr(\gamma^\alpha W^L)$.
Their complex conjugate operators $\overline T^{(\alpha)}_{L}$ carry opposite orbifold twist. 

If we build correlators from such BPS operators, we have to contract every $\phi_I$ field with a conjugate $\bar{\phi}_I$, which in the supergraph corresponds to simply connecting the respective operators by a vector-multiplet propagator. Since the twist operators $\gamma$ commute with the $V^{++}$ fields and the twist lines can likewise be moved through vector-multiplet propagators, the position of the twist lines in the supergraph is arbitrary and does not affect the value of the correlator. 

Quantum corrections will introduce internal gauge interactions and hypermultiplet loops, but as long as the latter do not encircle one of the external operators, their intersection with twist lines remains trivial. The first quantum correction with non-trivial twist-dependence arises when a hypermultiplet wraps the shortest twisted BPS operator of length $L_0$
\begin{equation}\label{eq: wrap part}
\begin{tikzpicture}[thick,baseline=-3pt]
\usetikzlibrary{decorations.pathmorphing}

\node[draw, circle, minimum size=1.0cm] (OL) at (0,0)
    {\scriptsize{$T_{L_0}^{(\alpha)}$}};

% Vertically aligned right endpoints
\coordinate (R1) at (2, 0.38);
\coordinate (R3) at (2, 0);
\coordinate (R4) at (2,-0.19);
\coordinate (R5) at (2,-0.38);

\draw[decorate,
      decoration={snake,aspect=1, amplitude=1pt, segment length=5.2pt}]
      (OL.45) -- (R1);

\draw[dashed,red] (OL.-45) -- (R5);

\draw[decorate,
      decoration={snake,aspect=1, amplitude=1pt, segment length=5.2pt}]
      (OL.-21) -- (R4);

\draw[decorate,
      decoration={snake,aspect=1, amplitude=1pt, segment length=5.2pt}]
      (OL.0) -- (R3);

% Vertical dots indicating more lines above and below
\node at (1.2,0.18)
    {$\vcenter{\hbox{\shortstack{$\cdot$\\[-5pt]$\cdot$\\[-5pt]$\cdot$}}}$};

% Horizontal dots indicating continuation to the right
\node at (2.35,0) {$\cdots$};

% Draw thicker circle on top
\draw[line width=1.5pt] (OL) circle [radius=0.5cm];

\draw[line width=1pt] (OL) ++(0.8cm,0)
arc[start angle=0,end angle=360,radius=0.8cm]
node[pos=0.125,sloped,allow upside down]
{\tikz{\draw[-{Triangle[length=5pt,width=5pt]}] (0,0)--(0.001,0);}};

\end{tikzpicture}
\quad\sim \omega^\alpha g^{2L_{0}}\,.
\end{equation}
This subdiagram\footnote{For simplicity, we drew effective quartic vertices here, which capture the combined effect of two cubic ones (see \cite{LongPaper} for details).} first starts contributing at $\order{g^{2L_{0}}}$, so all lower orders in perturbation theory match the untwisted case directly inherited from $\mathcal{N}=4$ SYM. Phrased in integrability language, we observe twist-dependence to arise only at \emph{wrapping order} in perturbation theory, while the \emph{asymptotic regime} is determined by the refined inheritance principle.

We will now discuss these twist-dependent wrapping corrections in detail for small numbers of $T^{(\alpha)}_L$ operators and see that hypermultiplet circles around twisted sector operators provide a natural organisation principle, reflected by both localisation and integrability results. The explicit evaluation of the supergraphs is performed in \cite{LongPaper} and the results are presented here.

\subsection{Two-point functions} \label{Sec:TwoPointFunction}

In order to isolate the effect of the orbifold twist, we may normalise the twisted correlators against their untwisted counterparts. Take for example the two-point function
\begin{equation}\label{eq: ratio}
    \mathcal{R}^{(\alpha)}_{L} \equiv \frac{\left\langle T^{(\alpha)}_{L}(x)\, \overline T^{(\alpha)}_{L}(y) \right\rangle}{\left\langle T^{(0)}_{L}(x)\,\overline T^{(0)}_{L}(y) \right\rangle}\,.
\end{equation}
The leading contribution to the numerator that is not cancelled by the denominator is given by the following  wrapping diagram at order $\order{g^{2L}}$

\begin{equation}\label{eq: 2pt}
\begin{array}{c c}
\begin{tikzpicture}[thick,baseline=-3pt]
\usetikzlibrary{decorations.pathmorphing}
\node[draw, circle, minimum size=1.0cm] (OL) at (0,0) {\scriptsize{$T^{(\alpha)}_L$}};
\node[draw, circle, minimum size=1.0cm] (OBar) at (2,0) {\scriptsize{$\overline T^{(\alpha)}_L$}};

\draw[decorate,
      decoration={snake,aspect=1, amplitude=1pt, segment length=5.2pt}]
      (OL.45) -- (OBar.135);

\draw[dashed,red] (OL.-45) -- (OBar.-135);

\draw[decorate,
      decoration={snake,aspect=1, amplitude=1pt, segment length=5.2pt}]
      (OL.-21) -- (OBar.-159);

\draw[decorate,
      decoration={snake,aspect=1, amplitude=1pt, segment length=5.2pt}]
      (OL.0) -- (OBar.180);

\node at (1,0.18)
    {$\vcenter{\hbox{\shortstack{$\cdot$\\[-5pt]$\cdot$\\[-5pt]$\cdot$}}}$};
% Draw thicker circles on top (no shifting!)
\draw[line width=1.5pt] (OL) circle [radius=0.5cm];
\draw[line width=1.5pt] (OBar) circle [radius=0.5cm];

\draw[line width=1pt] (OL) ++(0.8cm,0)
arc[start angle=0,end angle=360,radius=0.8cm]
node[pos=0.125,sloped,allow upside down]
{\tikz{\draw[-{Triangle[length=5pt,width=5pt]}] (0,0)--(0.001,0);}};
\end{tikzpicture}\,\, +\,\,  \begin{tikzpicture}[thick,baseline=-3pt]
\usetikzlibrary{decorations.pathmorphing}
\node[draw, circle, minimum size=1.0cm] (OL) at (0,0) {\scriptsize{$T^{(\alpha)}_L$}};
\node[draw, circle, minimum size=1.0cm] (OBar) at (2,0) {\scriptsize{$\overline T^{(\alpha)}_L$}};

\draw[decorate,
      decoration={snake,aspect=1, amplitude=1pt, segment length=5.2pt}]
      (OL.45) -- (OBar.135);

\draw[dashed,red] (OL.-45) -- (OBar.-135);

\draw[decorate,
      decoration={snake,aspect=1, amplitude=1pt, segment length=5.2pt}]
      (OL.-21) -- (OBar.-159);

\draw[decorate,
      decoration={snake,aspect=1, amplitude=1pt, segment length=5.2pt}]
      (OL.0) -- (OBar.180);

\node at (1,0.18)
    {$\vcenter{\hbox{\shortstack{$\cdot$\\[-5pt]$\cdot$\\[-5pt]$\cdot$}}}$};
% Draw thicker circles on top (no shifting!)
\draw[line width=1.5pt] (OL) circle [radius=0.5cm];
\draw[line width=1.5pt] (OBar) circle [radius=0.5cm];

\draw[line width=1pt] (OL) ++(0.8cm,0)
arc[start angle=0,end angle=360,radius=0.8cm]
node[pos=0.125,sloped,allow upside down]
{\tikz{\draw[-{Triangle[length=5pt,width=5pt]}] (0.001,0)--(0,0);}};
\end{tikzpicture}
\end{array}
\end{equation}
while the same diagrams without the twist lines appear in the numerator and thus have to be subtracted. In total we find the phase factor 
\begin{equation}
    \omega^\alpha+\omega^{-\alpha}-2=-4 s_\alpha\,, \quad s_\alpha\equiv\sin^2\frac{\pi \alpha}{M}\,.
\end{equation}
The full superspace evaluation of these diagrams reduces to a point-split ladder integral, which has been evaluated in \cite{Usyukina:1993ch} to the conformal period
\begin{equation}\label{eq: period}
	W_L=\binom{2L}{L}\zeta(2L-1)\,,
\end{equation}
and collecting these contributions we find
\begin{equation} \label{eq:TwoPointWrapping}
	\mathcal{R}_{L}^{(\alpha)}(g) = 1-4s_\alpha g^{2L} W_L + O(g^{2L+2})\,.
\end{equation}
This agrees exactly with the weak coupling expansion of localisation results  \cite{Beccaria:2022ypy,Billo:2022fnb,Billo:2022lrv,Korchemsky:2025eyc} and an $\mathcal{N}=1$ superspace calculation performed in \cite{Galvagno:2020cgq}.

The natural expectation is now that further quantum corrections to \eqref{eq: 2pt} will generate a power series in $g^2$ multiplying the same phase $s_\alpha$. Once we reach order $\order{g^{4L}}$, however, we can produce a second hypermultiplet circle wrapping the operator and so on. Thus an organisation by numbers of wrapped hypermultiplets leads to a natural double expansion
\begin{equation}\label{eq: two-point structure}
    \mathcal{R}^{(\alpha)}_{L}(g)=\sum_{n,m=0}^\infty a_{n,m}  s_\alpha^{n}g^{2Ln+2m}\,,
\end{equation}
with $a_{0,m}=\delta_{m,0}$ and $a_{n,0}=(-4W_L)^n $ \cite{LongPaper}. The same structure was observed in the localisation literature \cite{Korchemsky:2025eyc} hinting at a supergraph correspondence to be discussed below.

\subsection{Extremal three-point functions}
We now turn towards three-point functions of $T^{(\alpha)}_L$ operators \eqref{eq:twisted-operator}. Their only non-vanishing correlators (up to overall complex conjugation) are extremal correlators in which two operators $T^{(\alpha_1)}_{L_1}$, $T^{(\alpha_2)}_{L_2}$ are contracted completely with a conjugate operator $\overline T^{(\alpha_3)}_{L_3}$ satisfying $\alpha_3=\alpha_1+\alpha_2\text{ mod }M$ and $L_3=L_1+L_2$. The associated structure constant ratio is given by 
\begin{equation} \label{eq:Norm3ptFktExt}
	\mathcal{C}^{(\alpha_1,\alpha_2,\alpha_3)}_{L_1,L_2,L_3} := \frac{\left\langle T_{L_1}^{(\alpha_1)}(x_1)\, T_{L_2}^{(\alpha_2)}(x_2)\, \overline T_{L_3}^{(\alpha_3)}(x_3) \right\rangle}{\left\langle T_{L_1}^{(0)}(x_1)\, T_{L_2}^{(0)}(x_2)\, \overline T_{L_3}^{(0)}(x_3) \right\rangle}\,.
\end{equation}

The refined inheritance principle dictates that twist-dependence only appears once a virtual hypermultiplet circle closes around a twisted operator. We can identify three candidates for the leading wrapping correction:
\begin{equation*}
\begin{array}{c@{\quad\,\,}c@{\quad\,\,}c}

% ============================================================
% PANEL 1
% ============================================================
\begin{tikzpicture}[thick,baseline=-3pt,scale=0.6]

\tikzset{
  op/.style={
    draw=black,
    fill=white,
    circle,
    minimum size=0.54cm,
    inner sep=0pt,
    line width=1.2pt,
    font=\small
  },
  curl/.style={
    black,
    line width=0.8pt,
    decorate,
    decoration={
      snake,
      aspect=1,
      amplitude=0.75pt,
      segment length=5.0pt
    },
    shorten <=2.5mm,
    shorten >=2.5mm
  },
  contract/.style={
    red,
    line width=0.9pt,
    dash pattern=on 3pt off 2pt,
    shorten <=2.5mm
  },
  highlight/.style={
    draw=black,
    line width=0.8pt
  },
  banddots/.style={
    font=\normalsize,
    inner sep=0pt,
    text=black
  }
}

\coordinate (T1) at (0,1.62);
\coordinate (L1) at (-1.40,-0.81);
\coordinate (R1) at ( 1.40,-0.81);
\coordinate (C1) at (0,0);

% ------------------------------------------------------------
% Red dashed contractions
% ------------------------------------------------------------
\draw[contract] (T1) -- (C1);
\draw[contract] (L1) -- (C1);
\draw[contract] (R1) -- (C1);

% Red central vertex
\fill[red] (C1) circle (1.6pt);

% Large circle around lower-left operator
\draw[highlight] (L1) circle[radius=0.80cm];

% ------------------------------------------------------------
% Band: top -- lower left
% Innermost snake stays fixed
% ------------------------------------------------------------

% outermost
\draw[curl]
  ([shift={(-0.095, 0.055)}]T1) --
  ([shift={(-0.095, 0.055)}]L1);

% middle
\draw[curl]
  ([shift={(-0.015, 0.009)}]T1) --
  ([shift={(-0.015, 0.009)}]L1);

% innermost -- fixed
\draw[curl]
  ([shift={( 0.065,-0.038)}]T1) --
  ([shift={( 0.065,-0.038)}]L1);

% ------------------------------------------------------------
% Band: top -- lower right
% Innermost snake stays fixed
% ------------------------------------------------------------

% innermost -- fixed
\draw[curl]
  ([shift={(-0.065,-0.038)}]T1) --
  ([shift={(-0.065,-0.038)}]R1);

% middle
\draw[curl]
  ([shift={( 0.015, 0.009)}]T1) --
  ([shift={( 0.015, 0.009)}]R1);

% outermost
\draw[curl]
  ([shift={( 0.095, 0.055)}]T1) --
  ([shift={( 0.095, 0.055)}]R1);

% ------------------------------------------------------------
% Compact dots indicating additional lines
% Slightly farther from outermost snake
% ------------------------------------------------------------
\path
  ($(T1)!0.50!(L1)+(-0.336,0.194)$)
  node[banddots,rotate=150]
  {$\mathinner{\cdot\!\!\cdot\!\!\cdot}$};

\path
  ($(T1)!0.50!(R1)+(0.336,0.194)$)
  node[banddots,rotate=30]
  {$\mathinner{\cdot\!\!\cdot\!\!\cdot}$};

% ------------------------------------------------------------
% Operators
% ------------------------------------------------------------
\node[op] at (T1) {3};
\node[op] at (L1) {1};
\node[op] at (R1) {2};

\end{tikzpicture}

&

% ============================================================
% PANEL 2
% ============================================================
\begin{tikzpicture}[thick,baseline=-3pt,scale=0.6]

\tikzset{
  op/.style={
    draw=black,
    fill=white,
    circle,
    minimum size=0.54cm,
    inner sep=0pt,
    line width=1.2pt,
    font=\small
  },
  curl/.style={
    black,
    line width=0.8pt,
    decorate,
    decoration={
      snake,
      aspect=1,
      amplitude=0.75pt,
      segment length=5.0pt
    },
    shorten <=2.5mm,
    shorten >=2.5mm
  },
  contract/.style={
    red,
    line width=0.9pt,
    dash pattern=on 3pt off 2pt,
    shorten <=2.5mm
  },
  highlight/.style={
    draw=black,
    line width=0.8pt
  },
  banddots/.style={
    font=\normalsize,
    inner sep=0pt,
    text=black
  }
}

\coordinate (T2) at (0,1.62);
\coordinate (L2) at (-1.40,-0.81);
\coordinate (R2) at ( 1.40,-0.81);
\coordinate (C2) at (0,0);

% ------------------------------------------------------------
% Red dashed contractions
% ------------------------------------------------------------
\draw[contract] (T2) -- (C2);
\draw[contract] (L2) -- (C2);
\draw[contract] (R2) -- (C2);

% Red central vertex
\fill[red] (C2) circle (1.6pt);

% Large circle around lower-right operator
\draw[highlight] (R2) circle[radius=0.80cm];

% outermost
\draw[curl]
  ([shift={(-0.095, 0.055)}]T2) --
  ([shift={(-0.095, 0.055)}]L2);

% middle
\draw[curl]
  ([shift={(-0.015, 0.009)}]T2) --
  ([shift={(-0.015, 0.009)}]L2);

% innermost -- fixed
\draw[curl]
  ([shift={( 0.065,-0.038)}]T2) --
  ([shift={( 0.065,-0.038)}]L2);

% innermost -- fixed
\draw[curl]
  ([shift={(-0.065,-0.038)}]T2) --
  ([shift={(-0.065,-0.038)}]R2);

% middle
\draw[curl]
  ([shift={( 0.015, 0.009)}]T2) --
  ([shift={( 0.015, 0.009)}]R2);

% outermost
\draw[curl]
  ([shift={( 0.095, 0.055)}]T2) --
  ([shift={( 0.095, 0.055)}]R2);

\path
  ($(T2)!0.50!(L2)+(-0.336,0.194)$)
  node[banddots,rotate=150]
  {$\mathinner{\cdot\!\!\cdot\!\!\cdot}$};

\path
  ($(T2)!0.50!(R2)+(0.336,0.194)$)
  node[banddots,rotate=30]
  {$\mathinner{\cdot\!\!\cdot\!\!\cdot}$};

\node[op] at (T2) {3};
\node[op] at (L2) {1};
\node[op] at (R2) {2};

\end{tikzpicture}

&

\begin{tikzpicture}[thick,baseline=-3pt,scale=0.6]

\tikzset{
  op/.style={
    draw=black,
    fill=white,
    circle,
    minimum size=0.54cm,
    inner sep=0pt,
    line width=1.2pt,
    font=\small
  },
  curl/.style={
    black,
    line width=0.8pt,
    decorate,
    decoration={
      snake,
      aspect=1,
      amplitude=0.75pt,
      segment length=5.0pt
    },
    shorten <=2.5mm,
    shorten >=2.5mm
  },
  contract/.style={
    red,
    line width=0.9pt,
    dash pattern=on 3pt off 2pt,
    shorten <=2.5mm
  },
  highlight/.style={
    draw=black,
    line width=0.8pt
  },
  banddots/.style={
    font=\normalsize,
    inner sep=0pt,
    text=black
  }
}

\coordinate (T3) at (0,1.62);
\coordinate (L3) at (-1.40,-0.81);
\coordinate (R3) at ( 1.40,-0.81);
\coordinate (C3) at (0,0);

\draw[contract] (T3) -- (C3);
\draw[contract] (L3) -- (C3);
\draw[contract] (R3) -- (C3);

% Red central vertex
\fill[red] (C3) circle (1.6pt);

% Large circle around top operator
\draw[highlight] (T3) circle[radius=0.80cm];

% outermost
\draw[curl]
  ([shift={(-0.095, 0.055)}]T3) --
  ([shift={(-0.095, 0.055)}]L3);

% middle
\draw[curl]
  ([shift={(-0.015, 0.009)}]T3) --
  ([shift={(-0.015, 0.009)}]L3);

% innermost -- fixed
\draw[curl]
  ([shift={( 0.065,-0.038)}]T3) --
  ([shift={( 0.065,-0.038)}]L3);

% innermost -- fixed
\draw[curl]
  ([shift={(-0.065,-0.038)}]T3) --
  ([shift={(-0.065,-0.038)}]R3);

% middle
\draw[curl]
  ([shift={( 0.015, 0.009)}]T3) --
  ([shift={( 0.015, 0.009)}]R3);

% outermost
\draw[curl]
  ([shift={( 0.095, 0.055)}]T3) --
  ([shift={( 0.095, 0.055)}]R3);

\path
  ($(T3)!0.50!(L3)+(-0.336,0.194)$)
  node[banddots,rotate=150]
  {$\mathinner{\cdot\!\!\cdot\!\!\cdot}$};

\path
  ($(T3)!0.50!(R3)+(0.336,0.194)$)
  node[banddots,rotate=30]
  {$\mathinner{\cdot\!\!\cdot\!\!\cdot}$};

\node[op] at (T3) {3};
\node[op] at (L3) {1};
\node[op] at (R3) {2};

\end{tikzpicture}\,,

\end{array}
\end{equation*}
which may be evaluated to yield a total correction
\begin{equation}
    \label{eq: 3pt leading}
    \mathcal{C}^{(\alpha_1,\alpha_2,\alpha_3)}_{L_1,L_2,L_3} =1-4 \sum_{i=1}^3s_{\alpha_i}g^{2L_i}W_{L_i}+\dots\,.
\end{equation}
The contributions of circles around $T^{(\alpha_1)}_{L_1}$, $T^{(\alpha_2)}_{L_2}$ arise in the same way as for the two-point correlator. We confirmed explicitly that the contribution of the \emph{split circle} around $\overline T^{(\alpha_3)}_{L_3}$ reduces to $W_{L_3}$ \cite{LongPaper}.
As before, we expect higher-order quantum corrections to start contributing at $\mathcal{O}(g^{2\min(L_i)+2})$. A direct generalisation to multiple hypermultiplet circles suggests an organisation of the perturbative series into towers of twist factors $s_{\alpha_i}$
\begin{align}
    \mathcal{C}^{(\alpha_1,\alpha_2,\alpha_3)}_{L_1,L_2,L_3}&=\sum_{n_i=0}^\infty d_{n_1,n_2,n_3}(g)\,s_{\alpha_1}^{n_1}s_{\alpha_2}^{n_2}s_{\alpha_3}^{n_3}\,g^{2(\sum_in_iL_i)}\,,\nonumber\\ \label{eq: 3-pt}
    d_{n_1,n_2,n_3}(g)&=\prod_{i=1}^3(-4)^{n_i}W_{L_i}^{n_i} +\order{g^2}\,,
\end{align}
where the leading term in the $d_{n_1,n_2,n_3}$ coefficients factorises into the elementary wrapping contributions \eqref{eq: wrap part} of individual hypermultiplet circles as we show in \cite{LongPaper}. This factorisation has likewise been observed in localisation and integrability and allows a graph-level matching of contributions as we will discuss below.

In contrast to the two-point ratio $\mathcal{R}^{(\alpha)}_L$ \eqref{eq: ratio} which may be absorbed into the normalisation of twisted operators, the structure constant ratio $\mathcal{C}^{(\alpha_1,\alpha_2,\alpha_3)}_{L_1,L_2,L_3}$ can be made normalisation-independent by dividing it by $(\mathcal{R}^{(\alpha_1)}_{L_1}\mathcal{R}^{(\alpha_2)}_{L_2}\mathcal{R}^{(\alpha_3)}_{L_3})^{1/2}$. This combination then constitutes an honest physical observable of the theory.

\subsection{Non-extremal three-point functions}
An extension to non-extremal BPS three-point functions requires the introduction of additional BPS operators that feature a combination of $\phi_I$ and $\bar \phi_I$ fields. As discussed in \cite{Gadde:2009dj,Mantegazza:2026spd} such BPS operators do not occur in the twisted sector but can appear in the untwisted sector ($\alpha=0$). Diagonalisation of the dilatation operator suggests that the appropriate untwisted BPS operators are given by a particular combination of totally symmetrised traces
\begin{equation}
\tilde{U}_{K,L} \sim  \tr\left( \phi^{K}\bar \phi^{L} \right) - \frac{1}{2} \tr\left( \phi^{K-1} f \bar \phi^{L-1} \bar{f}\right)+\ldots \,,
\label{eq:BPS-first-components}
\end{equation}
where the dots denote further terms with additional insertions of hypermultiplet scalar pairs $f\bar{f}$ \cite{LongPaper}. Operators of this form are often referred to as \emph{vacuum descendants}, since they are related to $T^{(0)}_{L+K}$ via an R-symmetry transformation in the parent theory. We may then consider the non-extremal three-point function ratio 
\begin{equation}
\tilde{\mathcal{C}}^{(\alpha)}_{L_1,L_2,L_3} = \frac{\braket{ T_{L_1}^{(\alpha)}(x_1)\, \overline{T}_{L_2}^{(\alpha)}(x_2)\,\tilde{U}_{\ell_{23},\ell_{13}}(x_3) }}{\braket{ T_{L_1}^{(0)}(x_1)\, \overline{T}_{L_2}^{(0)}(x_2)\, \tilde{U}_{\ell_{23},\ell_{13}}(x_3) }}\,,
\label{eq:nonext-correlator}
\end{equation}
where $L_3=\ell_{13}+\ell_{23}$ and the bridge-lengths $\ell_{13}$, $\ell_{23}$ and 
\begin{equation}
    \ell_{12}=\frac{L_1+L_2-\ell_{13}-\ell_{23}}{2}\,,
\end{equation}
indicate how many vector-multiplet propagators connect the different operators at tree level.

The hypermultiplet pair $f\bar{f}$ in \eqref{eq:BPS-first-components} gives rise to a new kind of leading twist-sensitive graph. In addition to hypermultiplet circles fully enclosing twisted operators, we can form bifundamental circles that originate and terminate at the composite operator $\tilde{U}_{\ell_{23},\ell_{13}}(x_3)$, \eg
\begin{equation}\label{eq: punctured}
\begin{tikzpicture}[thick,baseline=3pt,scale=0.6]

\tikzset{
  op/.style={
    draw=black,
    fill=white,
    circle,
    minimum size=0.54cm,
    inner sep=0pt,
    line width=1.2pt,
    font=\small
  },
  curl/.style={
    black,
    line width=0.8pt,
    decorate,
    decoration={
      snake,
      aspect=1,
      amplitude=0.75pt,
      segment length=5.0pt
    },
    shorten <=2.5mm,
    shorten >=2.5mm
  },
  curlextra/.style={
    black,
    line width=0.8pt,
    decorate,
    decoration={
      snake,
      aspect=1,
      amplitude=0.75pt,
      segment length=5.0pt
    }
  },
  contract/.style={
    red,
    line width=0.9pt,
    dash pattern=on 3pt off 2pt
  },
  highlight/.style={
    draw=black,
    line width=0.8pt
  },
  banddots/.style={
    font=\normalsize,
    inner sep=0pt,
    text=black
  }
}

\coordinate (T1) at (0,1.62);
\coordinate (L1) at (-1.40,-0.81);
\coordinate (R1) at ( 1.40,-0.81);

\coordinate (DL) at ($(L1)+(30.052:0.45cm)$);
\coordinate (DR) at ($(R1)+(149.948:0.45cm)$);

\draw[contract] (DL) -- (DR);

\node[
  circle,
  minimum size=0.54cm,
  inner sep=0pt,
  outer sep=0pt
] (TopAnchor) at (T1) {};

% Geometry parameters
\def\wrapR{0.8}

% Controls the lower curved branch
\def\lowerOut{0.8}
\def\entryDepth{0.9}

\coordinate (WrapUpper)
  at ($(L1)+(150.0524:\wrapR cm)$);

\coordinate (WrapLower)
  at ($(L1)+(-29.9476:\wrapR cm)$);
  
% Exact lowest point of operator 3
\coordinate (BottomEnd) at (TopAnchor.south);

% Continue initially parallel to the 1--3 direction
\coordinate (LowerCtrl)
  at ($(WrapLower)+(60.0524:\lowerOut cm)$);

% Final control point directly below BottomEnd:
% therefore the curve enters operator 3 vertically
\coordinate (EntryCtrl)
  at ($(BottomEnd)+(0,-\entryDepth cm)$);

\def\topHitAngle{202}
\def\topEntryLength{0.72}

\coordinate (TopEnd)
  at ($(T1)+(\topHitAngle:0.45cm)$);

% Reflect BottomEnd
\coordinate (ProjEnd)
  at ($(L1)!(BottomEnd)!(T1)$);

% Reflect EntryCtrl
\coordinate (ProjEntry)
  at ($(L1)!(EntryCtrl)!(T1)$);

\coordinate (TopEntryCtrl)
  at ($(EntryCtrl)!2!(ProjEntry)$);

% Reflect LowerCtrl
\coordinate (ProjLowerCtrl)
  at ($(L1)!(LowerCtrl)!(T1)$);

\coordinate (UpperCtrl)
  at ($(LowerCtrl)!2!(ProjLowerCtrl)$);

\draw[highlight]

  % mirrored upper curved branch
  (TopEnd)
  .. controls (TopEntryCtrl) and (UpperCtrl) ..
  (WrapUpper)

  % semicircular part around operator 1
  arc[
    start angle=150.0524,
    end angle=330.0524,
    radius=\wrapR cm
  ]

  % lower curved branch
  .. controls (LowerCtrl) and (EntryCtrl) ..
  (BottomEnd);

% outermost
\draw[curl]
  ([shift={(-0.095, 0.055)}]T1) --
  ([shift={(-0.095, 0.055)}]L1);

% middle
\draw[curl]
  ([shift={(-0.015, 0.009)}]T1) --
  ([shift={(-0.015, 0.009)}]L1);

% innermost
\draw[curl]
  ([shift={( 0.065,-0.038)}]T1) --
  ([shift={( 0.065,-0.038)}]L1);

% innermost
\draw[curl]
  ([shift={(-0.065,-0.038)}]T1) --
  ([shift={(-0.065,-0.038)}]R1);

% middle
\draw[curl]
  ([shift={( 0.015, 0.009)}]T1) --
  ([shift={( 0.015, 0.009)}]R1);

% outermost
\draw[curl]
  ([shift={( 0.095, 0.055)}]T1) --
  ([shift={( 0.095, 0.055)}]R1);

% upper
\draw[curl]
  ([shift={(0, 0.092)}]L1) --
  ([shift={(0, 0.092)}]R1);

% middle
\draw[curl]
  (L1) -- (R1);

% lower
\draw[curl]
  ([shift={(0,-0.092)}]L1) --
  ([shift={(0,-0.092)}]R1);

\coordinate (EL) at ($(L1)+(0.340,0.29)$);
\coordinate (ER) at ($(R1)+(-0.380,0.27)$);

% Shape parameters
\def\archX{0.4}
\def\archY{0.295}

% One continuous, horizontally mirror-symmetric arch
\draw[
  black,
  line width=0.8pt,
  decorate,
  decoration={
    snake,
    aspect=1,
    amplitude=0.7pt,
    segment length=4.83pt
  }
]
  (EL)
  .. controls (-\archX,\archY) and (\archX,\archY) ..
  (ER);

% 3 -- 1
\path
  ($(T1)!0.50!(L1)+(-0.336,0.194)$)
  node[banddots,rotate=150]
  {$\mathinner{\cdot\!\!\cdot\!\!\cdot}$};

% 3 -- 2
\path
  ($(T1)!0.50!(R1)+(0.336,0.194)$)
  node[banddots,rotate=30]
  {$\mathinner{\cdot\!\!\cdot\!\!\cdot}$};

% 1 -- 2
\path
  ($(L1)!0.50!(R1)+(0,-0.388)$)
  node[banddots,rotate=90]
  {$\mathinner{\cdot\!\!\cdot\!\!\cdot}$};

\node[op] at (T1) {3};
\node[op] at (L1) {1};
\node[op] at (R1) {2};

\end{tikzpicture}
\qquad \sim s_\alpha g^{2(\ell_{12}+1)}\vspace{-0.5em}
\end{equation}
This \emph{punctured} circle around $ T_{L_1}^{(\alpha)}$ intersects $\ell_{12}+1$ vector-multiplet propagators, which generically makes it the leading contribution. Note that the additional vector-multiplet propagator between  $ T_{L_1}^{(\alpha)}$ and  $ \overline T_{L_2}^{(\alpha)}$ had to be introduced because one power of $\phi\bar\phi$ in $\tilde{U}_{\ell_{23},\ell_{13}}$ was exchanged for the $f\bar{f}$ pair. 

The harmonic superspace evaluation of this diagram is presented in \cite{LongPaper} and can again be expressed in terms of the conformal period \eqref{eq: period}
\begin{equation} \label{eq:Non-extremalStructureConstant}
    \tilde{\mathcal{C}}^{(\alpha)}_{L_1,L_2,L_3}  = 1+ 4 s_\alpha g^{2 (\ell_{12}+1)} W_{\ell_{12}+1} + O(g^{2(\ell_{12}+2)})\,.
\end{equation}
 Only if one of the twisted operators satisfies $L_i=\ell_{12}+1$ the associated closed wrapping circle competes with this punctured circle contribution \cite{LongPaper}.

This extension to non-extremal BPS three-point functions demonstrates the predictive power of the harmonic superspace formalism beyond the regime charted by localisation. To our knowledge, \eqref{eq:Non-extremalStructureConstant} has not been computed before. A further extension to non-BPS correlators will inevitably involve a larger variety of supergraphs, but their relative simplicity suggests that their evaluation may still be feasible. This would allow us to generate the required field theory data to establish and validate exact analytical tools for orbifold theories. 

\section{Correspondence with Localisation}  

The comparison with localisation results for extremal correlators constitutes an important consistency check for our formalism and at the same time allows for a supergraph interpretation of individual terms in the localisation formulae. Take for example the two-point function ratio $\mathcal{R}^{(\alpha)}_L$ \eqref{eq: ratio} for which localisation predicts an exact expression \cite{Korchemsky:2025eyc}
\begin{equation}\label{eq: loc result}
     \mathcal{R}_{L}^{(\alpha)}(g) =\sum_{k=0}^{\infty} s_\alpha^k \left[K_{L-1}^k(g)\right]_{11}\,,
\end{equation}
in terms of the $(11)$-component of a power of semi-infinite matrices $K_{L-1}(g)$ with entries
\begin{equation}\begin{split}\label{eq: loc basic}
    (K_{L-1})_{ij} &\sim \int_0^\infty \frac{\de x}{x} \frac{J_{2i-2+L}(4gx) J_{2j-2+L}(4gx) }{\sinh^2(x)}\,,\\
   &\sim\order{g^{2(L+i+j-2)}}
\end{split}\end{equation}
where $J_{n}(x)$ denotes the Bessel function of the first kind. This expansion in powers of $s_\alpha$ exactly mimics the structure \eqref{eq: two-point structure} we derived from iterated hypermultiplet circles. This suggests that every entry of the $K_{L-1}(g)$ matrix can be thought of as the contribution of a hypermultiplet circle. The single-circle process \eqref{eq: 2pt} and all its higher-order corrections should be captured by $(K_{L-1})_{11}$. Two-circle processes should be compared to the $s_\alpha^2$ term in \eqref{eq: loc result}
\begin{equation}
   (K_{L-1})_{11}(K_{L-1})_{11}+ \sum_{i=2}^\infty (K_{L-1})_{1i}(K_{L-1})_{i1}\,,
\end{equation}
where the first term is simply a factorised repetition of the single-circle process. The additional terms come with additional powers of $g^{4(i-1)}$ suggesting that they arise from supergraphs with $i':=2(i-1)$ additional gauge-propagators connecting the two circles. Extrapolating from these cases, one may conjecture the following identification of the individual matrix entries
\begin{equation}\label{eq: conjecture}
    s_\alpha(K_{L-1})_{ij} \quad\sim \quad
    \begin{tikzpicture}[thick,baseline=-3pt]
\usetikzlibrary{decorations.pathmorphing}

\node[draw, dashed,circle, minimum size=0.6cm] (OI) at (0,0)
    {};
\node[draw, dashed,circle, minimum size=1.8cm] (OO) at (0,0)
{};
\node[draw,circle, minimum size=1.2cm] (OC) at (0,0)
{};

\draw[decorate,
      decoration={snake,aspect=1, amplitude=1pt, segment length=5.2pt}]
      (OI.20) -- (OO.20);

\draw[dashed,red] (OI.-45) -- (OO.-45);

\draw[decorate,
      decoration={snake,aspect=1, amplitude=1pt, segment length=5.2pt}]
      (OI.-20) -- (OO.-20);

\draw[decorate,
      decoration={snake,aspect=1, amplitude=1pt, segment length=5.2pt}]
      (OI.0) -- (OO.0);

\draw[decorate,
      decoration={snake,aspect=1, amplitude=1pt, segment length=5.2pt}]
      (OI.-160) -- (OC.-160);
\draw[decorate,
      decoration={snake,aspect=1, amplitude=1pt, segment length=5.2pt}]
      (OI.-130) -- (OC.-130);

\draw[decorate,
      decoration={snake,aspect=1, amplitude=1pt, segment length=5.2pt}]
      (OC.160) -- (OO.160);
\draw[decorate,
      decoration={snake,aspect=1, amplitude=1pt, segment length=5.2pt}]
      (OC.140) -- (OO.140);

\draw[line width=1pt] (OL) ++(0.6cm,0)
arc[start angle=0,end angle=360,radius=0.6cm];

\node[right] at (OO.0) {\scriptsize{$L$}};

\node[xshift=-6pt,yshift=2pt] at (OO.150) {\scriptsize{$j'$}};

\node[xshift=5pt,yshift=3pt] at (OI.-145) {\scriptsize{$i'$}};

\node[red,xshift=5pt,yshift=-3pt] at (OO.-45) {\scriptsize{$\alpha$}};

\end{tikzpicture}
    \,\,\,+ \,\,\,\text{corr.}\,,
\end{equation}
where the quantum corrections in $(K_{L-1})_{ij}$ sum over local decorations within the annulus and the full matrix multiplication in \eqref{eq: loc result} arises by stacking all connectible combinations of such annuli onto each other. At the current stage, this conjecture is supported by combinatorics, power-counting and transcendentality. We pose a further consolidation via a perturbative analysis of higher-order corrections as an interesting challenge for future research.

Such investigations are further motivated by localisation results for extremal three-point functions, which exhibit the factorisation observed in \eqref{eq: 3-pt} at all orders in $g$:
\begin{align}\label{eq: 3pt loc}
    \mathcal{C}^{(\alpha_1,\alpha_2,\alpha_3)}_{L_1,L_2,L_3} &=\mathcal{V}^{(\alpha_1)}_{L_1}\mathcal{V}^{(\alpha_2)}_{L_2}\mathcal{V}^{(\alpha_3)}_{L_3}\,\\
    \mathcal{V}^{(\alpha)}_{L}&=\sqrt{\mathcal{R}_{L}^{(\alpha)}(g)+\frac{g}{2L}\partial_g\mathcal{R}_{L}^{(\alpha)}(g)}\,.
\end{align}
The leading contributions of $\mathcal{V}^{(\alpha)}_L$ reproduce exactly the single-circle result \eqref{eq: 3pt leading}. We can match further contributions featuring powers of $(K_{L-1})_{11}$ since they just correspond to successive insertions of additional hypermultiplet circles \eqref{eq: 3-pt}, which factorise naturally. What remains to show is how the complete factorisation manifests itself for higher-order corrections. This is not directly obvious due to potential processes like
\begin{equation}\label{eq: non-fact}
\begin{tikzpicture}[thick,baseline=3pt,scale=0.6]

\tikzset{
  op/.style={
    draw=black,
    fill=white,
    circle,
    minimum size=0.54cm,
    inner sep=0pt,
    line width=1.2pt,
    font=\small
  },
  curl/.style={
    black,
    line width=0.8pt,
    decorate,
    decoration={
      snake,
      aspect=1,
      amplitude=0.75pt,
      segment length=5.0pt
    },
    shorten <=2.5mm,
    shorten >=2.5mm
  },
  contract/.style={
    red,
    line width=0.9pt,
    dash pattern=on 3pt off 2pt,
    shorten <=2.5mm
  },
  highlight/.style={
    draw=black,
    line width=0.8pt
  },
  banddots/.style={
    font=\normalsize,
    inner sep=0pt,
    text=black
  }
}

\coordinate (T1) at (0,1.62);
\coordinate (L1) at (-1.40,-0.81);
\coordinate (R1) at ( 1.40,-0.81);
\coordinate (C1) at (0,0);

\draw[contract] (T1) -- (C1);
\draw[contract] (L1) -- (C1);
\draw[contract] (R1) -- (C1);

% Red central vertex
\fill[red] (C1) circle (1.6pt);

\draw[highlight] (L1) circle[radius=0.80cm];

% outermost
\draw[curl]
  ([shift={(-0.095, 0.055)}]T1) --
  ([shift={(-0.095, 0.055)}]L1);

% middle
\draw[curl]
  ([shift={(-0.015, 0.009)}]T1) --
  ([shift={(-0.015, 0.009)}]L1);

% innermost
\draw[curl]
  ([shift={( 0.065,-0.038)}]T1) --
  ([shift={( 0.065,-0.038)}]L1);

% innermost
\draw[curl]
  ([shift={(-0.065,-0.038)}]T1) --
  ([shift={(-0.065,-0.038)}]R1);

% middle
\draw[curl]
  ([shift={( 0.015, 0.009)}]T1) --
  ([shift={( 0.015, 0.009)}]R1);

% outermost
\draw[curl]
  ([shift={( 0.095, 0.055)}]T1) --
  ([shift={( 0.095, 0.055)}]R1);

\def\bridgeAngle{20}

% Start on the large circle
\coordinate (BridgeStart)
  at ($(L1)+(\bridgeAngle:0.80cm)$);

\coordinate (BridgeEnd)
  at ($(L1)+(\bridgeAngle:2.42cm)$);

\draw[
  black,
  line width=0.8pt,
  decorate,
  decoration={
    snake,
    aspect=1,
    amplitude=0.75pt,
    segment length=5.0pt
  }
]
  (BridgeStart) -- (BridgeEnd);

\path
  ($(T1)!0.50!(L1)+(-0.336,0.194)$)
  node[banddots,rotate=150]
  {$\mathinner{\cdot\!\!\cdot\!\!\cdot}$};

\path
  ($(T1)!0.50!(R1)+(0.336,0.194)$)
  node[banddots,rotate=30]
  {$\mathinner{\cdot\!\!\cdot\!\!\cdot}$};

\node[op] at (T1) {3};
\node[op] at (L1) {1};
\node[op] at (R1) {2};

\end{tikzpicture}
\end{equation}
connecting the propagator bands between all operators.

\section{Correspondence with Integrability} 
Integrability provides powerful methods for evaluating correlation functions such as the hexagon formalism \cite{Basso:2015zoa} and the currently developing Separation of Variables formalism \cite{Bargheer:2025kli,Bargheer:2026kon}. Building on intuition from twisted string correlators, we proposed a twisted hexagon formalism for $\mathcal{N}=2$ orbifold theories in \cite{lePlat:2025eod} and tested it at tree level in gauge theory.
In the Appendix we present further explicit checks at one-loop order which suggest that this formalism holds for asymptotically long operators. The key ingredient in the twisted hexagon formalism was the introduction of additional twist lines which mimic the ones we now introduced in harmonic superspace.

Once we consider finite-length operators, additional finite-size effects have to be taken into account, which are generated by virtual excitations propagating on the dual string worldsheet. Whenever these excitations cross twist lines, they may pick up twist factors. This striking parallel with the hypermultiplet circle corrections we discussed above suggests a concrete matching of hexagon finite-size effects to twist-dependent supergraphs.

In \cite{Ferrando:2025qkr}, a first proposal for finite-size effects in extremal BPS three-point functions was successfully matched with localisation results \eqref{eq: 3pt loc}. Applying the methods presented there to the correlators we discussed above, we observe the following supergraph matching:

\subsection{Bridge contributions}

The hexagon formalism cuts the three-point correlator into two hexagons by introducing a full set of virtual intermediate states on the cutting lines, which are built from virtual excitations called \emph{mirror magnons}. For each combination of intermediate states one needs to compute the appropriate hexagon form factors and ultimately sum over all combinations. Each magnon crosses the \emph{bridge} of $\ell$ propagators connecting its two neighbouring physical operators, resulting in at least a $g^{2\ell}$ suppression of the associated \emph{bridge contribution}.

We observe a direct analogy to the punctured circle in \eqref{eq: punctured} which we may now interpret as bridge contribution in the hexagon formalism.

\begin{equation}\label{eq: bridge}
\begin{tikzpicture}[thick,baseline=3pt,scale=0.6]

\tikzset{
  op/.style={
    draw=black,
    fill=white,
    circle,
    minimum size=0.54cm,
    inner sep=0pt,
    line width=1.2pt,
    font=\small
  },
  curl/.style={
    black,
    line width=0.8pt,
    decorate,
    decoration={
      snake,
      aspect=1,
      amplitude=0.75pt,
      segment length=5.0pt
    },
    shorten <=2.5mm,
    shorten >=2.5mm
  },
  curlextra/.style={
    black,
    line width=0.8pt,
    decorate,
    decoration={
      snake,
      aspect=1,
      amplitude=0.75pt,
      segment length=5.0pt
    }
  },
  contract/.style={
    red,
    line width=0.9pt,
    dash pattern=on 3pt off 2pt
  },
  highlight/.style={
    draw=black,
    line width=0.8pt
  },
  banddots/.style={
    font=\normalsize,
    inner sep=0pt,
    text=black
  }
}

\coordinate (T1) at (0,1.62);
\coordinate (L1) at (-1.40,-0.81);
\coordinate (R1) at ( 1.40,-0.81);

\coordinate (DL) at ($(L1)+(30.052:0.45cm)$);
\coordinate (DR) at ($(R1)+(149.948:0.45cm)$);

\draw[contract] (DL) -- (DR);

\node[
  circle,
  minimum size=0.54cm,
  inner sep=0pt,
  outer sep=0pt
] (TopAnchor) at (T1) {};

\def\wrapR{0.8}
\def\lowerOut{0.8}
\def\entryDepth{0.9}

\coordinate (WrapUpper)
  at ($(L1)+(150.0524:\wrapR cm)$);

\coordinate (WrapLower)
  at ($(L1)+(-29.9476:\wrapR cm)$);

% Lower branch
\coordinate (BottomEnd) at (TopAnchor.south);

\coordinate (LowerCtrl)
  at ($(WrapLower)+(60.0524:\lowerOut cm)$);

\coordinate (EntryCtrl)
  at ($(BottomEnd)+(0,-\entryDepth cm)$);

% Upper branch
\def\topHitAngle{202}
\def\topEntryLength{0.72}

\coordinate (TopEnd)
  at ($(T1)+(\topHitAngle:0.45cm)$);

\coordinate (ProjEnd)
  at ($(L1)!(BottomEnd)!(T1)$);

\coordinate (ProjEntry)
  at ($(L1)!(EntryCtrl)!(T1)$);

\coordinate (TopEntryCtrl)
  at ($(EntryCtrl)!2!(ProjEntry)$);

\coordinate (ProjLowerCtrl)
  at ($(L1)!(LowerCtrl)!(T1)$);

\coordinate (UpperCtrl)
  at ($(LowerCtrl)!2!(ProjLowerCtrl)$);

% Entire wrapping contour
\draw[highlight]
  (TopEnd)
  .. controls (TopEntryCtrl) and (UpperCtrl) ..
  (WrapUpper)
  arc[
    start angle=150.0524,
    end angle=330.0524,
    radius=\wrapR cm
  ]
  .. controls (LowerCtrl) and (EntryCtrl) ..
  (BottomEnd);

\draw[curl]
  ([shift={(-0.095, 0.055)}]T1) --
  ([shift={(-0.095, 0.055)}]L1);

\draw[curl]
  ([shift={(-0.015, 0.009)}]T1) --
  ([shift={(-0.015, 0.009)}]L1);

\draw[curl]
  ([shift={( 0.065,-0.038)}]T1) --
  ([shift={( 0.065,-0.038)}]L1);

\draw[curl]
  ([shift={(-0.065,-0.038)}]T1) --
  ([shift={(-0.065,-0.038)}]R1);

\draw[curl]
  ([shift={( 0.015, 0.009)}]T1) --
  ([shift={( 0.015, 0.009)}]R1);

\draw[curl]
  ([shift={( 0.095, 0.055)}]T1) --
  ([shift={( 0.095, 0.055)}]R1);

\draw[curl]
  ([shift={(0, 0.092)}]L1) --
  ([shift={(0, 0.092)}]R1);

\draw[curl]
  (L1) -- (R1);

\draw[curl]
  ([shift={(0,-0.092)}]L1) --
  ([shift={(0,-0.092)}]R1);

\coordinate (EL) at ($(L1)+(0.340,0.29)$);
\coordinate (ER) at ($(R1)+(-0.380,0.27)$);

\def\archX{0.4}
\def\archY{0.295}

\draw[
  black,
  line width=0.8pt,
  decorate,
  decoration={
    snake,
    aspect=1,
    amplitude=0.7pt,
    segment length=4.83pt
  }
]
  (EL)
  .. controls (-\archX,\archY) and (\archX,\archY) ..
  (ER);

\path
  ($(T1)!0.50!(L1)+(-0.336,0.194)$)
  node[banddots,rotate=150]
  {$\mathinner{\cdot\!\!\cdot\!\!\cdot}$};

\path
  ($(T1)!0.50!(R1)+(0.336,0.194)$)
  node[banddots,rotate=30]
  {$\mathinner{\cdot\!\!\cdot\!\!\cdot}$};

\path
  ($(L1)!0.50!(R1)+(0,-0.388)$)
  node[banddots,rotate=90]
  {$\mathinner{\cdot\!\!\cdot\!\!\cdot}$};

\node[op] at (T1) {3};
\node[op] at (L1) {1};
\node[op] at (R1) {2};

\end{tikzpicture}
\quad\Leftrightarrow\quad
\begin{tikzpicture}[thick,baseline=-3pt,scale=0.6]

\tikzset{
  hexsolid/.style={
    draw=black,
    line width=1.0pt,
    line cap=round
  },
  hexdashed/.style={
    draw=black,
    line width=1.0pt,
    dashed,
    line cap=round
  },
  topred/.style={
    draw=red,
    line width=1.0pt,
    line cap=round,
    dashed
  },
  bluesquare/.style={
    draw=blue!60!black,
    fill=white,
    rectangle,
    minimum size=0.20cm,
    inner sep=0pt,
    line width=0.9pt
  }
}

% Side length of each regular hexagon
\def\hexs{1.2}
\pgfmathsetmacro{\hexh}{0.8660254*\hexs}

\begin{scope}[xshift=0cm]

\coordinate (A1) at (-0.5*\hexs,-\hexh);
\coordinate (B1) at ( 0.5*\hexs,-\hexh);
\coordinate (C1) at ( \hexs,0);
\coordinate (D1) at ( 0.5*\hexs, \hexh);
\coordinate (E1) at (-0.5*\hexs, \hexh);
\coordinate (F1) at (-\hexs,0);

% Reflected hexagon outline
\draw[hexsolid]  (F1) -- (A1);
\draw[hexsolid]  (B1) -- (C1);
\draw[hexdashed] (C1) -- (D1);
\draw[hexsolid]  (D1) -- (E1);
\draw[hexdashed] (E1) -- (F1);

% Dashed edge now at the bottom
\draw[hexdashed] (A1) -- (B1);

\draw[topred]
  ($(A1)+(0,0.12)$) --
  ($(B1)+(0,0.12)$);

% Blue square now on bottom edge
\node[bluesquare]
  at ($(A1)!0.50!(B1)$) {};

\end{scope}

\node[font=\large] at (1.65,0) { $\,\times\,$ };

\begin{scope}[xshift=3.30cm]

\coordinate (A2) at (-0.5*\hexs,-\hexh);
\coordinate (B2) at ( 0.5*\hexs,-\hexh);
\coordinate (C2) at ( \hexs,0);
\coordinate (D2) at ( 0.5*\hexs, \hexh);
\coordinate (E2) at (-0.5*\hexs, \hexh);
\coordinate (F2) at (-\hexs,0);

% Reflected hexagon outline
\draw[hexsolid]  (F2) -- (A2);
\draw[hexsolid]  (B2) -- (C2);
\draw[hexdashed] (C2) -- (D2);
\draw[hexsolid]  (D2) -- (E2);
\draw[hexdashed] (E2) -- (F2);

% Dashed edge now at the bottom
\draw[hexdashed] (A2) -- (B2);

% Blue square now on bottom edge
\node[bluesquare]
  at ($(A2)!0.50!(B2)$) {};

\end{scope}

\end{tikzpicture}
\end{equation}

The leading finite-size correction is given by a single magnon on this bridge
\begin{equation}
    \delta \mathcal{B}^{(\alpha)}= \sum_{Q=1}^\infty \int \frac{\de u}{2\pi} e^{-\ell \tilde{E}_Q(u)} \mu_Q(u) \, 4Q s_\alpha\,.
    \label{eq:BridgeContribution}
\end{equation}
where $Q$ is a bound-state index, $\tilde{E}_Q(u)$ is the mirror energy and $\mu_Q(u)$ the associated measure \cite{Basso:2015zoa}. Their respective weak coupling expansions are given by 
\begin{equation}
\begin{aligned}
    e^{-\tilde{E}_Q(u)} &=\frac{g^2}{\frac{Q^2}{4}+u^2}+\order{g^4}\,, \\ \mu_Q(u) &=\frac{Q\,g^2}{(\frac{Q^2}{4}+u^2)^2}+\order{g^4}\,.
\end{aligned}
\end{equation}
The only one-particle mirror contribution is the process \eqref{eq: bridge}, which at leading order evaluates to
\begin{equation}
\begin{aligned}
    \delta \mathcal{B}^{(\alpha)} &= 4 i^{\ell_{12}} \, s_\alpha g^{2(\ell_{12}+1)} \times \\
    &\phantom{====} \sum_{Q=1}^\infty Q \, \partial_u^{\ell_{12}+2} \, \frac{(\frac{Q}{2}-i u)^{\ell_{12}+2}}{(\ell_{12}+1)!} + \order{g^{2(\ell_{12}+2)}} \\
    &= 4 s_\alpha g^{2(\ell_{12}+1)} W_{\ell_{12}+1} + \order{g^{2(\ell_{12}+2)}}\,.
\end{aligned}
\end{equation}
Assuming that $\ell_{13},\ell_{23}>1$, this perfectly reproduces our harmonic superspace result \eqref{eq:Non-extremalStructureConstant}.
In the integrability framework, the extra factor of $g^2$ stems from the measure $\mu_k(u)\sim O(g^2)$, while in the field theory calculation this factor arises due to the two additional propagators connecting to the hypermultiplet circle in \eqref{eq: punctured}.

\subsection{Wrapping contributions}

Additional \emph{wrapping contributions} arise when a pair of magnons on adjacent edges have the same rapidity $u$ and thus form a combined bound state. The evaluation of this process in the hexagon formalism, \cf \cite{Ferrando:2025qkr}, captures the wrapping corrections \eqref{eq: wrap part} for the twisted operators, \eg
\begin{equation}
\begin{tikzpicture}[thick,baseline=3pt,scale=0.6]

\tikzset{
  op/.style={
    draw=black,
    fill=white,
    circle,
    minimum size=0.54cm,
    inner sep=0pt,
    line width=1.2pt,
    font=\small
  },
  curl/.style={
    black,
    line width=0.8pt,
    decorate,
    decoration={
      snake,
      aspect=1,
      amplitude=0.75pt,
      segment length=5.0pt
    },
    shorten <=2.5mm,
    shorten >=2.5mm
  },
  contract/.style={
    red,
    line width=0.9pt,
    dash pattern=on 3pt off 2pt,
    shorten <=2.5mm
  },
  highlight/.style={
    draw=black,
    line width=0.8pt
  },
  banddots/.style={
    font=\normalsize,
    inner sep=0pt,
    text=black
  }
}

\coordinate (T1) at (0,1.62);
\coordinate (L1) at (-1.40,-0.81);
\coordinate (R1) at ( 1.40,-0.81);
\coordinate (C1) at (0,0);

\draw[contract] (T1) -- (C1);
\draw[contract] (L1) -- (C1);
\draw[contract] (R1) -- (C1);

% Red central vertex
\fill[red] (C1) circle (1.6pt);

% Large circle around lower-left operator
\draw[highlight] (L1) circle[radius=0.80cm];

% outermost
\draw[curl]
  ([shift={(-0.095, 0.055)}]T1) --
  ([shift={(-0.095, 0.055)}]L1);

% middle
\draw[curl]
  ([shift={(-0.015, 0.009)}]T1) --
  ([shift={(-0.015, 0.009)}]L1);

% innermost -- fixed
\draw[curl]
  ([shift={( 0.065,-0.038)}]T1) --
  ([shift={( 0.065,-0.038)}]L1);

% innermost -- fixed
\draw[curl]
  ([shift={(-0.065,-0.038)}]T1) --
  ([shift={(-0.065,-0.038)}]R1);

% middle
\draw[curl]
  ([shift={( 0.015, 0.009)}]T1) --
  ([shift={( 0.015, 0.009)}]R1);

% outermost
\draw[curl]
  ([shift={( 0.095, 0.055)}]T1) --
  ([shift={( 0.095, 0.055)}]R1);

\draw[curl]
  ([shift={(0, 0.092)}]L1) --
  ([shift={(0, 0.092)}]R1);

\draw[curl]
  (L1) -- (R1);

\draw[curl]
  ([shift={(0,-0.092)}]L1) --
  ([shift={(0,-0.092)}]R1);

\path
  ($(L1)!0.50!(R1)+(0,-0.388)$)
  node[banddots,rotate=90]
  {$\mathinner{\cdot\!\!\cdot\!\!\cdot}$};

\path
  ($(T1)!0.50!(L1)+(-0.336,0.194)$)
  node[banddots,rotate=150]
  {$\mathinner{\cdot\!\!\cdot\!\!\cdot}$};

\path
  ($(T1)!0.50!(R1)+(0.336,0.194)$)
  node[banddots,rotate=30]
  {$\mathinner{\cdot\!\!\cdot\!\!\cdot}$};

\node[op] at (T1) {3};
\node[op] at (L1) {1};
\node[op] at (R1) {2};

\end{tikzpicture}
\quad\Leftrightarrow\quad
\begin{tikzpicture}[thick,baseline=-3pt,scale=0.6]

\tikzset{
  hexsolid/.style={
    draw=black,
    line width=1.0pt,
    line cap=round
  },
  hexdashed/.style={
    draw=black,
    line width=1.0pt,
    dashed,
    line cap=round
  },
  topred/.style={
    draw=red,
    line width=1.0pt,
    line cap=round,
    dashed
  },
  bluesquare/.style={
    draw=blue!60!black,
    fill=white,
    rectangle,
    minimum size=0.20cm,
    inner sep=0pt,
    line width=0.9pt
  }
}

% Side length of each regular hexagon
\def\hexs{1.2}
\pgfmathsetmacro{\hexh}{0.8660254*\hexs}

\begin{scope}[xshift=0cm]

\coordinate (A1) at (-0.5*\hexs,-\hexh);
\coordinate (B1) at ( 0.5*\hexs,-\hexh);
\coordinate (C1) at ( \hexs,0);
\coordinate (D1) at ( 0.5*\hexs, \hexh);
\coordinate (E1) at (-0.5*\hexs, \hexh);
\coordinate (F1) at (-\hexs,0);

\draw[topred]
  ($(A1)+(0,0.12)$) --
  ($(B1)+(0,0.12)$);
\draw[topred]
  ($(E1)+(0.11,-0.06)$) --
  ($(F1)+(0.11,-0.06)$);  
\draw[topred]
  ($(D1)+(-0.11,-0.06)$) --
  ($(C1)+(-0.11,-0.06)$);  

% Reflected hexagon outline
\draw[hexsolid]  (F1) -- (A1);
\draw[hexsolid]  (B1) -- (C1);
\draw[hexdashed] (C1) -- (D1);
\draw[hexsolid]  (D1) -- (E1);
\draw[hexdashed] (E1) -- (F1);

% Dashed edge now at the bottom
\draw[hexdashed] (A1) -- (B1);

% Blue square now on bottom edge
\node[bluesquare]
  at ($(A1)!0.50!(B1)$) {};
\node[bluesquare]
  at ($(E1)!0.50!(F1)$) {};

\end{scope}

\node[font=\large] at (1.65,0) { $\,\times\,$ };

\begin{scope}[xshift=3.30cm]

\coordinate (A2) at (-0.5*\hexs,-\hexh);
\coordinate (B2) at ( 0.5*\hexs,-\hexh);
\coordinate (C2) at ( \hexs,0);
\coordinate (D2) at ( 0.5*\hexs, \hexh);
\coordinate (E2) at (-0.5*\hexs, \hexh);
\coordinate (F2) at (-\hexs,0);

% Reflected hexagon outline
\draw[hexsolid]  (F2) -- (A2);
\draw[hexsolid]  (B2) -- (C2);
\draw[hexdashed] (C2) -- (D2);
\draw[hexsolid]  (D2) -- (E2);
\draw[hexdashed] (E2) -- (F2);

% Dashed edge now at the bottom
\draw[hexdashed] (A2) -- (B2);

% Blue square now on bottom edge
\node[bluesquare]
  at ($(A2)!0.50!(B2)$) {};
\node[bluesquare]
    at ($(C2)!0.50!(D2)$) {};

\end{scope}

\end{tikzpicture}
\end{equation}

The evaluation of the leading-order wrapping processes in the hexagon formalism, with one magnon on each adjacent edge of a twisted operator of length $L_1$ can be carried out as in \cite{Ferrando:2025qkr} and reads
\begin{equation} \label{eq:HexWrapping}
\begin{aligned}
    \delta \mathcal{W}^{(\alpha_1)} &= \frac{1}{2} \sum_{a=1}^\infty \int \frac{\de u}{2\pi} e^{-L_1 \tilde{E}_a(u)} \amsmathbb{W}^{(\alpha_1)} \\
   &= -2 s_{\alpha_1} g^{2 L_1} W_{L_1} +O(g^{2(L_1+1)})\,,
\end{aligned}
\end{equation}
where the relevant hexagon contact term $\amsmathbb{W}^{(\alpha_1)}$ was given in \cite{Ferrando:2025qkr}. We again observe a leading-order match with the harmonic superspace result \eqref{eq: 3pt leading} once we normalise the latter by $(\mathcal{R}^{(\alpha_1)}_{L_1}\mathcal{R}^{(\alpha_2)}_{L_2}\mathcal{R}^{(\alpha_3)}_{L_3})^{-1/2}$, as we discussed above.

\subsection{Beyond leading order}

We identified the leading supergraph counterparts to bridge and wrapping corrections. Subleading corrections on both field-theory and hexagon sides could be considered next in an attempt to expand this correspondence. An expansion to higher numbers of hypermultiplet circles is expected to correspond to an expansion to higher numbers of magnons.

The hexagon formalism \cite{Ferrando:2025qkr} predicts further so-called \emph{bridge-like contributions} which arise from a combination of magnons on all three bridges
\begin{equation}\label{eq: bridgelike}
\begin{tikzpicture}[thick,baseline=-3pt,scale=0.6]

\tikzset{
  hexsolid/.style={
    draw=black,
    line width=1.0pt,
    line cap=round
  },
  hexdashed/.style={
    draw=black,
    line width=1.0pt,
    dashed,
    line cap=round
  },
  topred/.style={
    draw=red,
    line width=1.0pt,
    line cap=round,
    dashed
  },
  bluesquare/.style={
    draw=blue!60!black,
    fill=white,
    rectangle,
    minimum size=0.20cm,
    inner sep=0pt,
    line width=0.9pt
  }
}

% Side length of each regular hexagon
\def\hexs{1.2}
\pgfmathsetmacro{\hexh}{0.8660254*\hexs}

\begin{scope}[xshift=0cm]

\coordinate (A1) at (-0.5*\hexs,-\hexh);
\coordinate (B1) at ( 0.5*\hexs,-\hexh);
\coordinate (C1) at ( \hexs,0);
\coordinate (D1) at ( 0.5*\hexs, \hexh);
\coordinate (E1) at (-0.5*\hexs, \hexh);
\coordinate (F1) at (-\hexs,0);

\draw[topred]
  ($(A1)+(0,0.12)$) --
  ($(B1)+(0,0.12)$);
\draw[topred]
  ($(E1)+(0.11,-0.06)$) --
  ($(F1)+(0.11,-0.06)$);  
\draw[topred]
  ($(D1)+(-0.11,-0.06)$) --
  ($(C1)+(-0.11,-0.06)$);  

% Reflected hexagon outline
\draw[hexsolid]  (F1) -- (A1);
\draw[hexsolid]  (B1) -- (C1);
\draw[hexdashed] (C1) -- (D1);
\draw[hexsolid]  (D1) -- (E1);
\draw[hexdashed] (E1) -- (F1);

% Dashed edge now at the bottom
\draw[hexdashed] (A1) -- (B1);

% Blue square now on bottom edge
\node[bluesquare]
  at ($(A1)!0.50!(B1)$) {};
\node[bluesquare]
  at ($(E1)!0.50!(F1)$) {};
\node[bluesquare]
  at ($(D1)!0.50!(C1)$) {};

\end{scope}

\node[font=\large] at (1.65,0) { $\,\times\,$ };

\begin{scope}[xshift=3.30cm]

\coordinate (A2) at (-0.5*\hexs,-\hexh);
\coordinate (B2) at ( 0.5*\hexs,-\hexh);
\coordinate (C2) at ( \hexs,0);
\coordinate (D2) at ( 0.5*\hexs, \hexh);
\coordinate (E2) at (-0.5*\hexs, \hexh);
\coordinate (F2) at (-\hexs,0);

% Reflected hexagon outline
\draw[hexsolid]  (F2) -- (A2);
\draw[hexsolid]  (B2) -- (C2);
\draw[hexdashed] (C2) -- (D2);
\draw[hexsolid]  (D2) -- (E2);
\draw[hexdashed] (E2) -- (F2);

% Dashed edge now at the bottom
\draw[hexdashed] (A2) -- (B2);

% Blue square now on bottom edge
\node[bluesquare]
  at ($(A2)!0.50!(B2)$) {};
\node[bluesquare]
    at ($(C2)!0.50!(D2)$) {};
\node[bluesquare]
  at ($(E2)!0.50!(F2)$) {};

\end{scope}

\end{tikzpicture}\,.
\end{equation}
These contributions are more suppressed than bridge and wrapping contributions but were crucial in \cite{Ferrando:2025qkr} to reproduce the factorised structure of the localisation result \eqref{eq: 3pt loc} for extremal three-point functions. In more generic scenarios, it remains an open question how twists are assigned to such processes and whether a matching to supergraphs is possible. Candidate supergraphs connecting all three bridges can be constructed perturbatively as in \eqref{eq: non-fact} and through more complicated punctured hypermultiplet contours.

An evaluation of such higher-order effects in harmonic superspace will generate the necessary field-theory data from which to reconstruct finite-size effects in the twisted hexagon formalism. Once we have extracted the correct construction pattern and twist attribution, integrability will provide access to non-BPS correlators and establish full control of $\mathcal{N}=2$ orbifold structure constants. 

\section{Discussion}

We have introduced a particularly simple formulation of $\mathcal{N}=2$ orbifold theories in harmonic superspace, in which planar supergraphs are essentially reduced to $\mathcal{N}=4$ SYM processes. The only difference consists in topological twist lines which act on hypermultiplet propagators they intersect. This construction makes the orbifold inheritance principle manifest at supergraph level.

We used this formalism to determine the leading twist-dependent corrections to a large class of BPS two- and three-point functions (see \eqref{eq:TwoPointWrapping}, \eqref{eq: 3pt leading} and \eqref{eq:Non-extremalStructureConstant}).
These are generated by hypermultiplet circles around twisted operators and are therefore highly suppressed. Our results agree with localisation, where the latter is applicable, but extend beyond that domain to non-extremal correlators. These initial results demonstrate the computational power of harmonic superspace, opening up a feasible route towards higher-order corrections, higher-point functions and non-BPS correlators.

The particularly simple structure of localisation results for extremal correlators led us to conjecture a direct matching of hypermultiplet circles to individual terms $(K_{L-1})_{ij}$ \eqref{eq: conjecture} in the exact localisation formulas. Challenging this conjecture is a worthwhile exercise in harmonic superspace manipulations which will simultaneously generate relevant data and provide insight into the factorisation mechanism observed in \eqref{eq: 3pt loc}.

Finally, we compared the hypermultiplet circles to bridge and wrapping corrections in the orbifold theory and again observed a matching of individual contributions. This is largely due to the presence of the twist factors $s_\alpha$ which allow for a distinction of individual contributions and therefore make $\mathcal{N}=2$ orbifold theories a controlled yet non-trivial testing ground for integrability. This distinguishing effect of twist factors was previously used in the quantum spectral curve and Separation of Variables literature to resolve degeneracies in the basic $Q$-functions underlying these methods \cite{Kazakov:2015efa,Gromov:2022waj,Bargheer:2025kli}.

An unresolved issue is the treatment of twist lines in the hexagon formalism. Note, for example, that \eqref{eq: bridge} assumed a twist line along the lower bridge of the hexagon, but the associated supergraph allows for topological deformations of the twist line. If we were to move the twist line away from the lower bridge, it is unclear that \eqref{eq:BridgeContribution} still applies. Similarly the twist assignment in \eqref{eq: bridgelike} is not directly obvious. Such ambiguities also apply to the results of \cite{Ferrando:2025qkr} and need to be resolved in order to complete the twisted hexagon formalism for arbitrary three-point functions. Harmonic superspace can provide the necessary gauge-theory data to inform this resolution.

We conclude with a few final observations:
\begin{itemize}
    \item The refined orbifold inheritance principle is similar in spirit to Filk's theorem which applies to the non-commutative $\beta$-deformation \cite{Filk:1996dm,Fokken:2013mza}.
    \item The class of supergraphs built from hypermultiplet circles \eqref{eq: wrap part} is reminiscent of the wheel diagrams arising in the fishnet model \cite{Gurdogan:2015csr,Gromov:2017cja,Ekhammar:2025fvn}. It would be interesting to apply the integrability methods developed for the fishnet model to this subsector of operators. 
    \item The operators $T^{(\alpha)}_L$ defined in \eqref{eq:twisted-operator} may in general contain $1/N$-suppressed multi-trace admixtures with the same quantum numbers, which for extremal correlators may nevertheless contribute at leading order in large-$N$. We have not performed a full orthogonalisation but instead restricted ourselves to the single-trace part as is done in localisation \cite{Billo:2022fnb}. This is also natural from the integrability perspective \cite{Basso:2015zoa,Ferrando:2025qkr}, since the standard hexagon formalism computes the single-trace part of correlation functions. Multi-trace admixtures have to be incorporated separately~\cite{Eden:2017ozn}.
\end{itemize}

\section*{Acknowledgements}
We thank Zoltan Bajnok, Burkhard Eden, and all participants of the workshop \emph{Let's twist again}, held in June 2026 at DESY Hamburg, for discussions on many points related to this work.
This work was supported by the German Research Foundation DFG under Germany’s Excellence Strategy – EXC 2121 ‘Quantum Universe’ (390833306) through the 2025 QU Postdoc Grant. DlP is grateful for funding through a Feodor Lynen Fellowship by the Alexander von Humboldt Foundation and further acknowledges the grant NKKP Advanced 152467.

\appendix

%\begin{comment}
\section{Asymptotic three-point functions} \label{App:Asymptotic}

We claim that the hexagon formalism proposed in \cite{lePlat:2025eod} holds asymptotically. Here, we present some explicit checks at one-loop order. The results are checked against field theory calculations using the \emph{Hamiltonian insertion method} \cite{Alday:2005nd}, where the one-loop Hamiltonian is inserted at the splitting points of the operators or equivalent spin chains and the two-loop eigenstate needs to be used.

In the orbifold theory, the twist lines $\gamma^\alpha$ are inserted between the operators. Consequently, twist factors $\omega^{\pm \alpha(p-q)}$ are picked up, whenever a propagator crosses the twist line \cite{Cavaglia:2020hdb}, depending on the charges $p$ and $q$ of the vacuum and the excitations. 
Moreover, it is worth stressing that the result is \emph{independent} of the distribution of the twist lines on the correlator, which is expected from the topological invariance of the twist.

Some explicit $\amsmathbb{Z}_2$ orbifold one-loop results are given in Table~\ref{tab:Z2CorrelatorsYYB}. The operators $\mathcal{B}^{\alpha,L}$ are of length $L$, have twist $\alpha$ and carry two longitudinal excitations. There are two possible eigenstates which we label by a subscript $\pm$ \cite{lePlat:2025eod}.

\begin{table}[h]
\centering
\resizebox{0.5\textwidth}{!}{%
\begin{tabular}{c||c|c|c}
Correlators & $L=6$ & $L=8$ & $L=10$\\
\hline \hline
$\braket{\mathcal{B}_{\pm}^{1,8} \hat{\mathcal{B}}_{\pm}^{1,8} \mathcal{O}_L}$ & $-\frac{1}{\sqrt{6}}+4\sqrt{\frac{2}{3}}g^2$ & $\sqrt{2}-10\sqrt{2}g^2$ & $-\frac{\sqrt{5}}{3\sqrt{2}}+\frac{4\sqrt{10}}{3}g^2$ \\
$\braket{\mathcal{B}_{\pm}^{1,8} \hat{\mathcal{B}}_{\mp}^{1,8} \mathcal{O}_L}$ & $\frac{1}{\sqrt{6}}-4\sqrt{\frac{2}{3}}g^2$ & $\sqrt{2}-10\sqrt{2}g^2$ & $\frac{\sqrt{5}}{3\sqrt{2}}-\frac{4\sqrt{10}}{3}g^2$ \\
\hline
$\braket{\mathcal{B}_{\pm}^{1,10} \hat{\mathcal{B}}_{\pm}^{1,8} \mathcal{O}_L}$ & $0-\sqrt{\frac{3}{2}}g^2$ & $0$ & $-\sqrt{\frac{5}{2}}+5\sqrt{10}g^2$ \\
$\braket{\mathcal{B}_{\pm}^{1,10} \hat{\mathcal{B}}_{\mp}^{1,8} \mathcal{O}_L}$ & $0+\sqrt{\frac{3}{2}}g^2$ & $0$ & $\sqrt{\frac{5}{2}}-5\sqrt{10}g^2$ \\
\hline
$\braket{\mathcal{B}_{\pm}^{1,10} \hat{\mathcal{B}}_{\pm}^{1,10} \mathcal{O}_L}$ & $-\frac{\sqrt{3}}{2\sqrt{2}}+2\sqrt{6}g^2$ & $\sqrt{2}-10\sqrt{2}g^2$ & $0$ \\
$\braket{\mathcal{B}_{\pm}^{1,10} \hat{\mathcal{B}}_{\mp}^{1,10} \mathcal{O}_L}$ & $\frac{\sqrt{3}}{2\sqrt{2}}-2\sqrt{6}g^2$ & $-\sqrt{2}+10\sqrt{2}g^2$ & $0$
\end{tabular}}
\caption{Asymptotic three-point functions at one-loop. The twisted operators $\mathcal{B}_{\pm}^{1,L'}$ carry two excitations $Y$, while the twisted operators $\hat{\mathcal{B}}_{\pm}^{1,L'}$ carry two excitations $\bar{Y}$. The untwisted vacuum $\mathcal{O}_L$ is of length $L$. The correlators vanish for $L<6$.} \label{tab:Z2CorrelatorsYYB}
\end{table}

The asymptotic three-point functions were also evaluated by integrability as in \cite{lePlat:2025eod} using the solutions to the asymptotic Bethe equations
\begin{equation}
	e^{i p_j L} \prod_{j=1,j\neq i} ^{M} S_{j,i} = \omega^{\alpha(p-q)} \,, \qquad e^{iP} = \prod_{j=1}^{M} e^{i p_j}= \omega^{\alpha p} \,, \label{eq:BetheEq}
\end{equation}
where the Zhukovsky variables $e^{i p_j}=\frac{x^+(u_j)}{x^-(u_j)}$ and the all-loop S~matrix \cite{Beisert:2005tm} are used. The charges $(p,q)$ depend on the choice of vacuum and excitations, \textit{cf.} \cite{Ideguchi:2004wm,lePlat:2025eod}. We tested all the $\SU(2)$ sectors and found perfect agreement with field theory results. In particular, Table \ref{tab:Z2CorrelatorsYYB} is reproduced for longitudinal excitations.

%\end{comment}

\bibliographystyle{apsref}
\bibliography{refs}

\end{document}